\documentclass[useAMS,usenatbib,usegraphicx,useepsfig]{mn2e}

\usepackage[usenames]{color}

\usepackage{amsmath}

\newcommand{\msun}{${\rm M_{\sun}}$}

\def\ltsima{$\; \buildrel < \over \sim \;$}
\def\simlt{\lower.5ex\hbox{\ltsima}}
\def\gtsima{$\; \buildrel > \over \sim \;$}
\def\simgt{\lower.5ex\hbox{\gtsima}}
\def\kpc{{\rm\,kpc}}

\def\msun{{\rm\,M_\odot}}


\def\deg{^\circ}

\def\ltsima{$\; \buildrel < \over \sim \;$}
\def\gtsima{$\; \buildrel > \over \sim \;$}

\title[The intrinsic ellipticity of dwarf spheroidal galaxies]{The intrinsic ellipticity of dwarf spheroidal galaxies:\\ constraints from the Andromeda system}
\author[J.-B. Salomon, R. A. Ibata, N. F. Martin and B. Famaey]{J.-B. Salomon$^{1}$\thanks{E-mail:
jean-baptiste.salomon@astro.unistra.fr}, R. A. Ibata$^{1}$, N. F. Martin$^{1}$ and B. Famaey$^{1}$\\
$^{1}$Observatoire astronomique de Strasbourg, Universit\'{e} de Strasbourg, CNRS, UMR 7550, 11 rue de l'Universit\'{e}, F-67000 Strasbourg, France}
\begin{document}

\date{Accepted 2015 March 31. Received 2015 March 31; in original form 2015 January 20}

\pagerange{\pageref{firstpage}--\pageref{lastpage}} \pubyear{2015}

\maketitle

\label{firstpage}

\begin{abstract}
We present a study of the intrinsic deprojected ellipticity distribution of the satellite dwarf galaxies of the Andromeda galaxy, assuming that their visible components have a prolate shape, which is a natural outcome of simulations. Different possibilities for the orientation of the major axis of the prolate dwarf galaxies are tested, pointing either as close as possible to the radial direction towards the centre of Andromeda, or tangential to the radial direction, or with a random angle in the plane that contains the major axis and the observer. We find that the mean intrinsic axis ratio is $\sim 1/2$, with small differences depending on the assumed orientation of the population. Our deprojections also suggest that a significant fraction of the satellites, $\sim 10$\%, are tidally disrupted remnants. We find that there is no evidence of any obvious difference in the morphology and major axis orientation between satellites that belong to the vast thin plane of co-rotating galaxies around Andromeda and those that do not belong to this structure.
\end{abstract}


\section{Introduction}
The Local Group can be thought of as the closest laboratory for testing cosmological predictions on small scales, including the dynamical formation and evolutionary history of galaxies \citep{Freeman02}.  In this context the dwarf spheroidal (dSph) galaxies are of particular interest, as they represent the smallest of galactic systems, and are thought to be embedded in primordial dark matter subhalos, thus giving us insights on the earliest epochs of galaxy formation in the Universe. The investigation of these satellites is thus one of the most active areas of Local Group research at present \citep{Tolstoy09}. Their morphological parameters, their masses, their spatial distribution around their host, or the number of observed satellites, can be confronted with predictions of the $\Lambda$ Cold Dark Matter ($\Lambda$CDM) model \citep{Kravtsov04, Tollerud14}. For example, their spatial distribution can provide insights on the shape of the dark matter halo of the host galaxy \citep{Penarrubia02}, whilst the dwarf galaxies morphological characteristics can provide insights on their formation history \citep{Zolotov11, Carraro14}. 

The very nature of the dSphs as faint, poorly populated systems, poses several observational challenges, even for their identification in the Local Group. Generally, they have been discovered through the identification of small overdensities of stars over the contaminating foreground stellar populations (and background population of mis-identified compact galaxies). Improvements in instrumentation and analysis techniques have led to the discovery of several new Local Group dSph galaxies in recent years  \citep{Belokurov06, Zucker06, Walsh07, McConnachie08, Martin09, Martin13}. The known population of dSphs in the Local Group has thereby increased considerably \citep{McConnachie12}, which is now beginning to allow us to draw statistically firm conclusions about the properties of the dSph population. For instance, a recent discovery has shown that half of the dwarf galaxy satellites of the Andromeda galaxy (M31) reside in a vast thin plane around their host (VTP), and appear to have coherent motions \citep{Ibata13}. This observational dynamical alignment is still not fully understood within the $\Lambda$CDM framework where the subhalos that might host dSphs are a priori expected to possess only a weak alignment \citep{Shaya13, Ibata14b, Pawlowski14b, Bahl14}. Finding whether the satellites that partake in the alignment have a particular intrinsic shape and orientation that distinguishes them from the other dSphs orbiting around M31 may help in assessing whether the two sub-populations have experienced a different dynamical history.

The morphological properties of satellite dwarf galaxies can also provide tests of theoretical frameworks of galaxy formation on small scales. These galaxies are usually described as a baryonic component embedded in a massive dark matter halo \citep{Walker09}, meaning that their stars are relatively protected from the tidal field of their giant host. But the formation of a plane of satellites (as apparently required in M31), or any past interactions could have produced strong tidal effects that could have strongly affected their baryonic distribution and make them partly lose their dark matter halo. Another possibility is that some dSphs galaxies are actually not embedded in dark matter halos at all, but are rather remnants of so-called tidal dwarf galaxies which would have been created in an encounter between two major galaxies in the distant past, a scenario which might explain the formation of thin planes of galaxies such as the one observed around Andromeda \citep{Pawlowski11, Hammer13} and around the Milky Way (e.g., \citealt{Kroupa05, Kroupa10, Pawlowski12}). Such tidal dwarf galaxies would be mostly devoid of dark matter \citep{Kroupa97}, and would presumably be much more prone to tidal deformation than those embedded in primordial dark matter halos. Some of the problems of this picture are in principle alleviated if the Modified Newtonian Dynamics framework \citep{Milgrom83, Famaey2012, Zhao13} is adopted \citep{Kroupa10,Pawlowski14a}. Some other solutions consider the possibility of dissipative dark matter, as for example, in \cite{Foot13} or in \cite{Randall14}.

In cosmological simulations, subhalos are generally rounder than the host halos. Subhalos that have masses that are large enough to host dwarf galaxies have axis ratios : $0.65 \leq c/a \sim b/a \leq 0.95$ \citep{VeraCiro14}. Observed dSphs are observed to be approximately roughly round: for the Milky Way, $\epsilon = 1 - a/b \sim 0.3$ \citep{Martin08}. But we know that this observed ellipticity is just a projected view, and the third dimension could thus hide more elongated shapes. However, deprojecting to recover the intrinsic ellipticity is a challenge because distance uncertainties to different sides of the same dSph are currently far too large to allow any differential depth measurement except maybe in the Magellanic Clouds \citep{Moretti14, Klein14}. Therefore, in this paper, we propose a method which allows us to probe the third dimension, and we focus ourselves on M31\footnote{The method presented here could be modified later on to be applied to dwarf satellites of the Milky Way.}.

In section~\ref{method} we summarize the general method employed in this contribution, justify our hypotheses and present our data. Section~\ref{code} presents all dwarf galaxies parameters used in the analysis, which have been extracted from the observations. In section~\ref{deprojection}, we define how our deprojection procedure is built. Results are then discussed in sections~\ref{results} and \ref{individual}, and our conclusions are drawn in section~\ref{conclusion}.

\section{Method}\label{method}
\subsection{Hypotheses}

\begin{figure*}
\begin{center}
\includegraphics[scale=0.35]{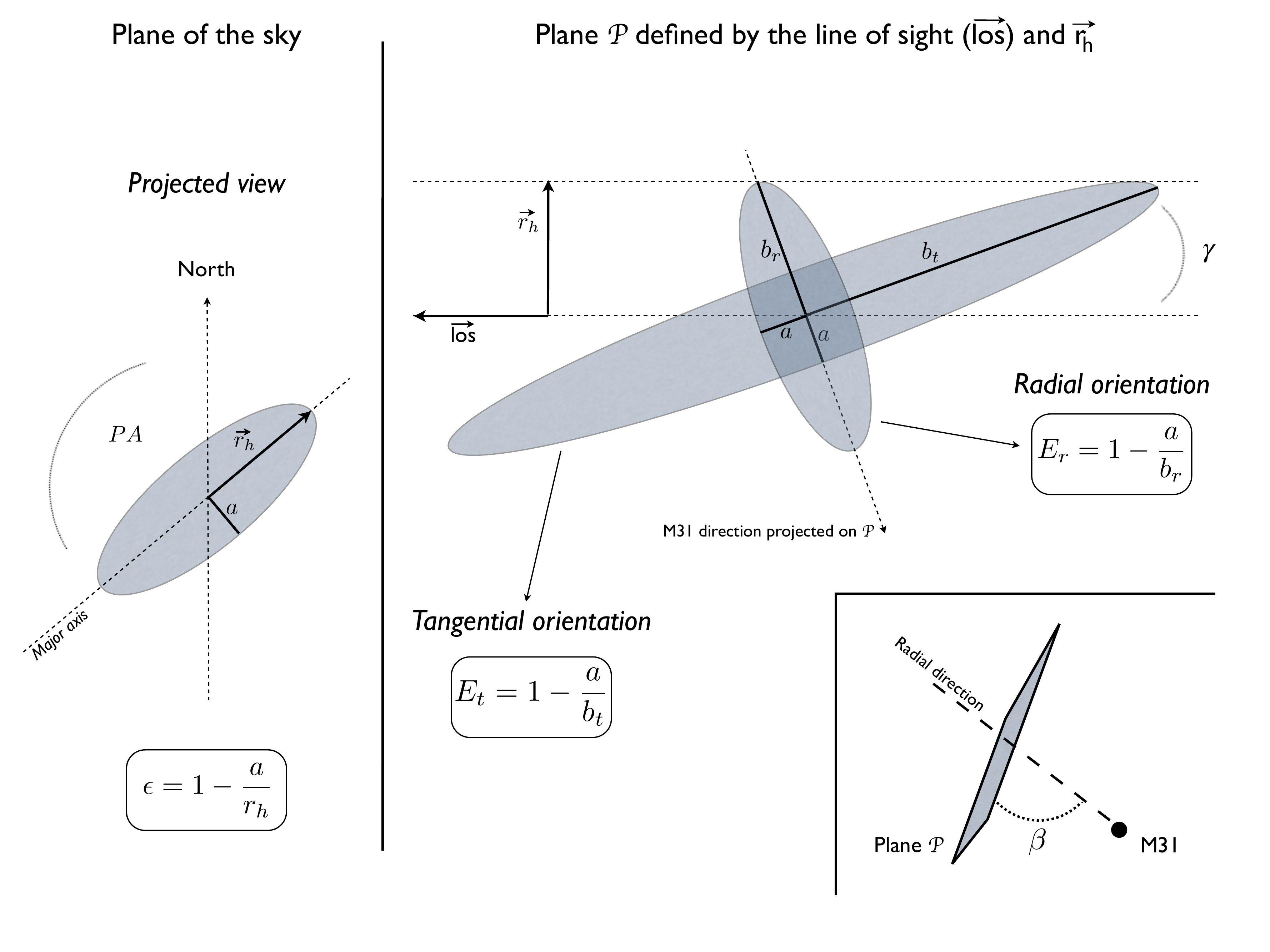}
\caption{\small\label{3dview}
Sketch of the geometry. The left-hand panel shows the projected view on the sky, while the right-hand panel shows the view in the plane $\cal{P}$ defined by the line of sight and the projected semimajor axis. Two cases are represented : the tangential case and the radial one. The bottom right-hand panel shows the position of the plane $\cal{P}$ compared to M31 where $\beta$ is the minimum angle between the radial direction and $\cal{P}$. The variable $a$ is the semi-minor axis of the prolate ellipsoid, equal to the semi-minor axis of the projected ellipse. On the left-hand panel, $PA$ is the angle between North and the semi-major axis $\vec{r_{h}}$ where $||\vec{r_{h}}||$ is the observed half-light radius. On the right-hand panel, $b_r$ is the intrinsic semi-major axis of the prolate ellipsoid placed in the configuration that is most radially-oriented with respect to M31, and $b_t$ is the one placed in the tangential configuration. The variable $\gamma$ is the angle between the major axis of the prolate ellipsoid and its intrinsic major axis. The line of sight position vector $\vec{los}$ links the Earth to the centre of the satellite galaxy.}
\end{center}
\end{figure*}

In order to deproject the view and try to see which intrinsic ellipticity distribution can fit the observed projected distribution of stellar populations in the target dwarf, some simplifying hypotheses need to be made. In the following, two main hypotheses are adopted. First, we consider that dSphs have a prolate shape, like a cigar. This is indeed the natural shape of dark matter halos in the relevant mass range \citep{VeraCiro14, Barber15}. In simulations in which dwarf satellites galaxies are initially made of rotating stellar discs, this is a less natural assumption, but they can in principle also be deformed into a prolate shape through bar-like instabilities caused by ongoing small tidal interactions with the host \citep{Lokas10}. Our second assumption will be to take an orientation of the prolate body that is consistent with the observed projection of the major axis, as sketched in Figure~\ref{3dview}. Three cases will be considered: 
\begin{enumerate}
\item
the prolate dwarf galaxy is oriented perpendicular to the radial vector linking the host and the satellite
\item
the major axis of the prolate dwarf galaxy is oriented as close as possible to the radial vector
\item
the prolate dwarf galaxy has a random orientation that is consistent with the observed projection.
\end{enumerate}

\subsection{Data}

In the present contribution, we will examine the structural properties of the dwarf galaxies surrounding the Andromeda galaxy (M31). Our observational data are derived from the Pan-Andromeda Archaeological Survey (PAndAS) \citep{McConnachie09}. This photometric survey with the MegaCam camera on the Canada-France-Hawaii Telescope has mapped a large area around Andromeda ($\approx$ 400 deg$^2$), allowing the discovery and characterization of the dwarf satellite galaxies of M31. A detailed account of the processing of the PAndAS survey through to the final catalogue of aperture magnitudes is given in \citep{Ibata14a}.

The PAndAS photometry was also used to estimate the distances to the dwarf galaxies, using the tip of the red giant branch (TRGB) method \citep{Conn12}. By using a Markov Chain Monte Carlo (MCMC) approach that incorporated Bayesian priors on the expected morphology of the satellites and their luminosity function, \citet{Conn12} were able to derive a posterior probability distribution function (PDF) for the distance to each of the satellites present in the survey.

From the homogeneous PAndAS data set, we also inferred the morphological parameters of all dwarf galaxies that fall in the survey footprint. These were derived using a version of the \cite{Martin08} algorithm, updated for a full MCMC treatment, and which will be presented in another publication that is to be submitted soon (Martin et al. 2015, in preparation).

\section{Extracting dwarf galaxy characteristics from the observations}\label{code}

\subsection{Morphology and selection}
The code developed by \cite{Martin08} infers the following observational parameters : the projected ellipticity ($\epsilon$, with $\epsilon = 1 - (a/b)$ where $a$ is the minor axis of the adjusted ellipse and $b$ the semimajor axis), the position angle ($PA$, the angle between the equatorial North and the major axis), and the half-light radius ($r_{h}$, taken along the major axis). The code was run on 23 dwarf galaxies present in the PAndAS survey and outputs posterior PDFs in the form of MCMC chains. We have also included in our study the dwarf elliptical satellites, NGC 147 and NGC 185. For them, realistic MCMC chains have been created by simultaneous drawings from gaussian PDFs built from \cite{Crnojevic14}, also derived from the PAndAS survey. According to \citet{Ibata13}, among these 25 satellites, 14 are in the VTP: And I, And III, And IX, And XI, And XII, And XIII, And XIV, And XVI, And XVII, And XXV, And XXVI, Cass II, NGC 147 and NGC 185. Whereas 11 are outside the VTP: And II, And V, And X, And XV, And XVIII, And XIX, And XX, And XXI, And XXII, And XXIII and And XXIV.

\begin{figure}
\begin{center}
\includegraphics[scale=0.32]{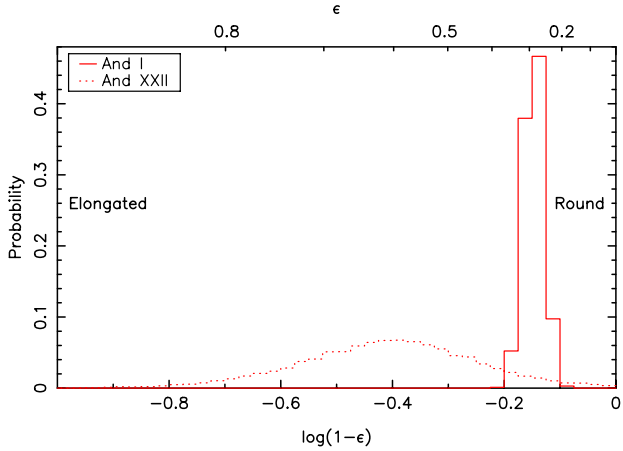}
\caption{\small\label{Nico}
Representative projected ellipticity distributions.
Histogram of the projected ellipticity for And I (solid line) and And XXII (dotted line) (two representative dSph in the sample) obtained from $10^5$ draws from the Markov chain of Martin et al. (2015) (in preparation).}
\end{center}
\end{figure}

Thus, for each satellite $k$ of Andromeda, the code provides us with a Markov chain that contains at each step $i$ the positional and morphological parameters of interest: $\epsilon_{ki}$, $PA_{ki}$ and $r_{{h}_{ki}}$.  For example, in Figure~\ref{Nico} we display the PDF derived from $10^5$ draws (this is typically the number of draws in all MCMC chains used in this study) of the projected ellipticity from the Markov chain of Andromeda I, an example of a rather round dSph galaxy with a large number of resolved red giant stars and Andromeda XXII, an example of an extended dSph galaxy with a small number of resolved red giant stars. It is more representative to plot $log(1 - \epsilon) = log(a/b)$ where $0$ corresponds to a round object.

\begin{figure}
\begin{center}
\includegraphics[scale=0.32]{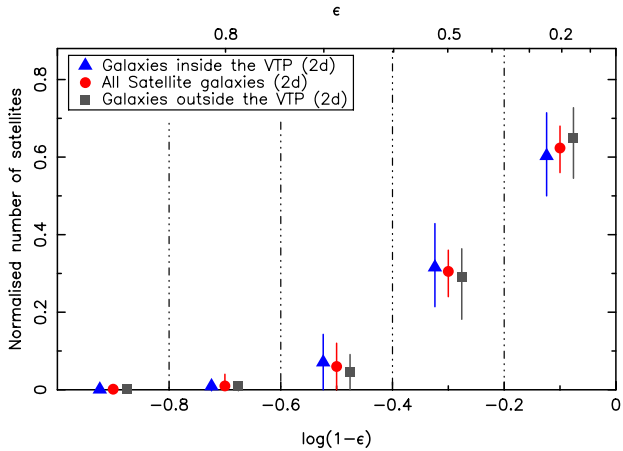}
\caption{\small\label{ProjectedEllipticity}
Observed (i.e., projected) ellipticity derived from the PAndAS survey, showing the average distribution of ellipticity for the 25 satellites, randomly drawn $10^5$ times from each of the 25 Markov chains. Bin sizes are $log(1 - \epsilon)=0.2$. Red circles are the ellipticity distribution of the entire system of 25 dwarf galaxies. Blue triangles correspond to the 14 dwarf galaxies contained in the vast thin plane of satellites \citep{Ibata13}, while the grey squares are for the 11 outside of the VTP. Vertical error bars indicate one standard deviation derived from this experiment.}
\end{center}
\end{figure}

When the whole sample of dSphs considered is taken into account, one draw from their respective Markov chain of ellipticity and distance represents a possible view from the Earth of the system of satellites around Andromeda. Figure~\ref{ProjectedEllipticity} represents the normalised average of the projected ellipticity of the satellites according to the derived PDFs for $10^5$ random realisations (of the 25 satellites). Bins have been chosen to be large enough, $log(1 - \epsilon)=0.2$, to ensure reliable statistics. The vertical bars are one sigma standard deviations of the number of satellites per bin, derived from  $7\times 10^4$ draws. 

As expected, the projected ellipticity shows predominantly round objects. Moreover, no significant difference between galaxies in the VTP (blue triangles) or outside it  (grey squares) is evident.

\subsection{Position in three dimensions}

\begin{figure}
\begin{center}
\includegraphics[scale=0.32]{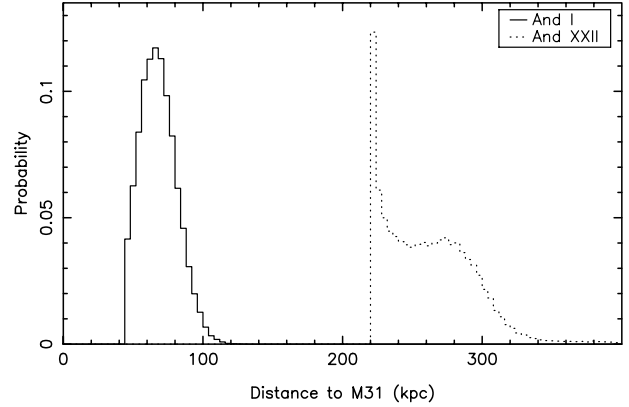}
\caption{\small\label{Anto}
Histogram of the distance of the centre of two representative systems, And I (solid line) and And XXII (dotted line) from the centre of M31 (derived from $10^5$ draws from the Markov chain of \citealt{Conn12}).}
\end{center}
\end{figure}

Once morphological parameters have been obtained, the position of each dSph centre in three dimensions has to be found. For that purpose, Andromeda's satellite galactic coordinates ($\alpha_{k}$, $\delta_{k}$) are considered perfect. We have chosen to take them from the literature in \cite{Collins13} for Cassiopeia II and in \citet{McConnachie12} for the other ones.
For the third dimension, the distance from Earth ($d$), we use results found by
\citet{Conn12} (see Figure~\ref{Anto}). This provides us with a Markov chain of distance for each galaxy $d_{k}$. Each step $i$ of it gives a possible $d_{ki}$ for the satellite.

For a dwarf galaxy, a drawing in its morphological Markov chain and one in its distance Markov chain combined with sky coordinates provide a set of six parameters. They define a possible representation in space of the centre around Andromeda in association with three possible projected morphological characteristics : $\alpha_{k}$, $\delta_{k}$, $d_{ki}$, $\epsilon_{ki}$, $PA_{ki}$ and $r_{{h}_{ki}}$.

\subsection{Model}

In all the following analysis, the intrinsic shape of the dwarf galaxies is approximated as a prolate ellipsoid, which is a natural outcome of dark matter satellites in $\Lambda$CDM simulations \citep{VeraCiro14, Barber15}. With this assumption the satellites have two equal minor axes ($a$) and one major axis ($b$). Consequently, since a prolate structure is symmetrical by rotation around the major axis, the projected minor axis remains identical, irrespective of the point of view. Thus the apparent minor axis of the ellipse given by the projected satellite is taken to measure the real minor axis ($a$). On the other hand, $\vec{r_{h}}$ is known, defined by $PA$ and $r_h$. 

Of course, due to the projection on the sky, the real major axis $\vec{b}$ is not equal to $\vec{r_{h}}$: $\vec{b}$ can have any orientation in the plane ($\cal{P}$) defined by the line of sight and $\vec{r_{h}}$ (see Figure~\ref{3dview}). Thus, an additional condition is required to fix the three-dimensional orientation of the major axis, and consequently the intrinsic ellipticity of the satellites. We consider three possibilities which will be discussed in Section~\ref{deprojection}.

\subsection{Limitation of the major axis}

As defined above, the intrinsic major axis $\vec{b}$ has to lie in the plane $\cal{P}$. However, depending on the parameter set drawn from the chain, it is possible that $b$ can take on physically implausible values. To avoid this situation, we implement a filter to reject solutions where $b$ would be larger than the tidal radius of the satellite (tidal rejection criterion). 
Thus, the tidal radius is considered as the maximal radius :
\begin{equation}\label{tidal}
b < \Big(\frac{M_{\rm sat}}{2 M_{\rm host}} \Big)^{1/3} R
\end{equation}
where $M_{\rm sat}$ is the satellite mass and $M_{\rm host}$ the mass of the host galaxy contained in the sphere of radius $R$ (distance between the satellite centre and the host galaxy centre). We estimate $M_{\rm host}$ using an NFW \citep{NFW97} halo model (with virial mass $2\times10^{12} \msun$, a mean value of recent results \citep{Watkins10, Fardal13}). The value of $M_{\rm sat}$ is derived from the line of sight velocity dispersion ($\sigma_{v}$) listed in \citet{McConnachie12} for NGC 147 and NGC 185 and in \cite{Collins13} for the other galaxies, taking:
\begin{equation}\label{disp}
	M_{\rm sat} \sim {\frac{\sigma_{v}^{2} \, b}{G}}
\end{equation}
where $G$ is the gravitational constant. Consequently, with equation~(\ref{tidal}) and~(\ref{disp}), the following limit  can be set to $b$ :
\begin{equation}\label{max}
	b < \sigma_{v}\sqrt{\frac{R^{3}}{2 \, G \, M_{\rm host}}}
\end{equation}
Given the large uncertainties in $\sigma_{v}$, we wish to remain conservative and so as not to eliminate physically possible cases, we adopt the values corresponding to the listed value of $\sigma_{v}$ plus one standard deviation. In the following analysis, this criterion is always automatically applied, and it forces the rejection of approximately 5\% of the MCMC parameter trial values. Finally, to avoid the MCMC process from exploring unrealistic solutions, we limited the projected half-light radius to 20 arcmin.

\subsection{Simple test}

\begin{figure}
\begin{center}
\includegraphics[scale=0.32]{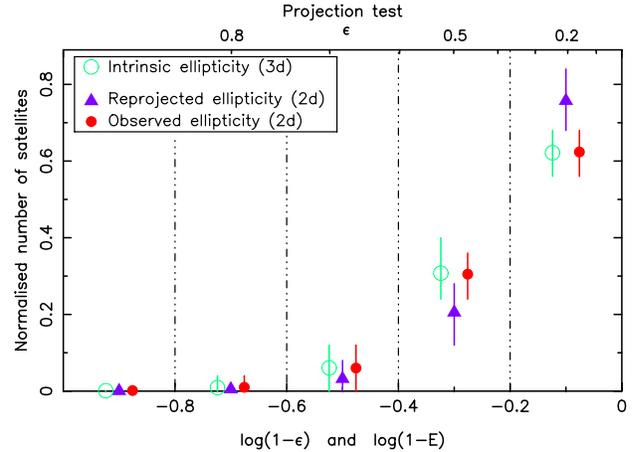}
\caption{\small\label{obsEll_reproj}
Green circles represent the intrinsic ellipticity chosen for this test. Purple triangles represent the distribution of projected ellipticity for 1000 3-dimensional systems of 25 dSph galaxies generated from a random distribution of position and from the green distribution for the ellipticity. Red dots are the observed ellipticities previously seen in Figure~\ref{ProjectedEllipticity}. Vertical error bars indicate one standard deviation.}
\end{center}
\end{figure}

We constructed a simple toy model to test the algorithms that we developed and to obtain a first assessment of the magnitude of the effect of de-projection on the ellipticity distribution. Three-dimensional prolate dwarf galaxies are generated with a random size (in the range of the observed dwarf satellites galaxies) and are placed at random locations around the Andromeda galaxy (again in the observed range). In Figure~\ref{obsEll_reproj} we show the consequence of assuming that the intrinsic ellipticity is given by the observed {\em projected} distribution previously displayed in Figure~\ref{ProjectedEllipticity} (red dots). These prolate structures are observed from a position corresponding to Earth, and we measure the projected ellipticity that the artificial structures would then be seen to possess (purple triangles). As expected, the projected distribution will appear rounder than the intrinsic distribution. The differences in the rounder bin is about 2$\sigma$, thus confirming that the projected distribution cannot be seen as the 3-dimensional one.

\section{Deprojection}\label{deprojection}

Our next task is to find the distribution of intrinsic ellipticities that is consistent with the observed projected distribution of Figure~\ref{ProjectedEllipticity}, and with the observed orientations and distances of the real satellites. As stated above, this will require an additional geometric constraint, which we will implement via an assumption regarding the 3-dimensional orientation of the satellites.

\subsection{Tangential deprojection}

As a first hypothesis, we suppose that each satellite is oriented tangentially with respect to the vector $\vec{r_s}$ connecting it to M31, i.e. that the real semimajor axis ($b$) of the modeled dwarf galaxy is on the plane $\cal{P'}$ perpendicular to $\vec{r_s}$. For satellite $k$, a given set $i$ of parameters ($l_{k}$, $b_{k}$, $d_{ki}$) then uniquely defines the direction of the vector $\vec{b}$, which corresponds to the intersection of $\cal{P}$ and $\cal{P'}$. Once the direction of $\vec{b}$ is known, $||\vec{b}||$ is determined by requiring that it correspond to the projected view of the dwarf galaxy whose apparent semimajor axis is $||\vec{r_{h}}||$. Finally, the intrinsic ellipticity $E_{T_{ki}}$ assuming this ``tangential hypothesis'', is calculated (see Figure~\ref{Tang}). In Appendix~\ref{annexetangent} we describe this deprojection in more detail.

\begin{figure}
\begin{center}
\includegraphics[scale=0.32]{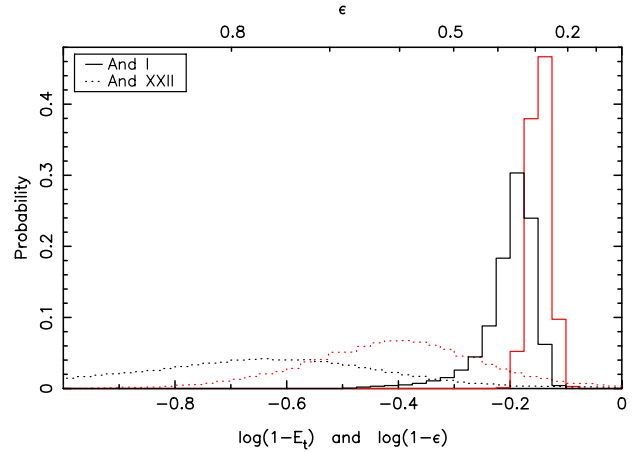}
\caption{\small\label{Tang}
Histogram of the ellipticity for And I (solid black line) and And XXII (dotted black line) after tangential deprojection obtained from $10^5$ draws of the MCMC chain. The red curves are the observed ellipticity previously seen in Figure~\ref{Nico}.}
\end{center}
\end{figure}

\begin{figure}
\begin{center}
\includegraphics[scale=0.32]{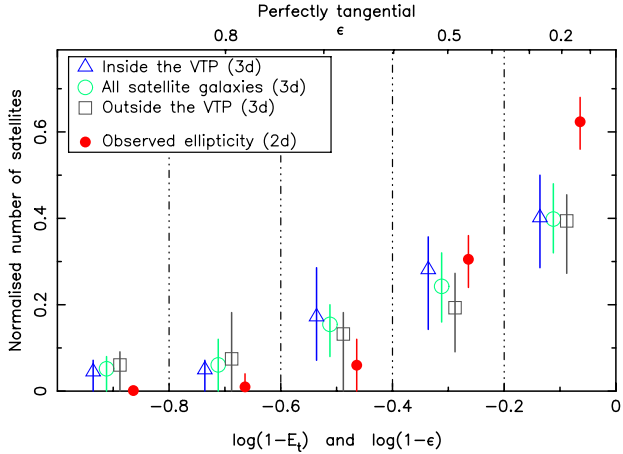}
\caption{\small\label{all_tang}
Distribution of tangentially deprojected ellipticity for $10^5$ systems generated from Markov chains derived from the observational analysis. Green circles mark the ellipticity distribution of the entire system of 25 dwarf galaxies. Blue triangles are for those contained in the vast thin plane of galaxies \citep{Ibata13}, 14 dSphs and grey squares are for those outside the VTP, 11 dSphs. Red dots are the observed ellipticities previously seen in Figure~\ref{ProjectedEllipticity}. Vertical error bars indicate one standard deviation.}
\end{center}
\end{figure}

As the tangential deprojected chain of each of the dwarf galaxies is now known, a random draw from each of the Markov chains gives a system of satellites all of which are in tangential orientation. This procedure is repeated to obtain the global ellipticity distribution of the dwarf galaxy system, which is shown in Figure~\ref{all_tang}. Under this ``tangential'' assumption, the intrinsic distribution of ellipticity is substantially flatter than the projected ellipticity, and produces a long tail of highly flattened systems. No significant difference between galaxies within the VTP or outside of it is apparent.

\subsection{Radial deprojection}

The second hypothesis that we explore in this analysis is to consider that the satellites are oriented as close as possible to the radial direction with respect to Andromeda. As before, the major axis $\vec{b}$ must lie on $\cal{P}$. For this ``radial'' deprojection, we would like $\vec{b}$ to lie along $\vec{r_s}$. However, the prolate ellipsoid must also be consistent with the observed (projected) orientation of the galaxy. In this situation, we have three spatial unknowns that have to respect four independent equations, so that the problem turns out to be over-constrained. Thus, an exact radial orientation is, in general, impossible. 

A way to resolve the problem is to attempt to orient the prolate ellipsoid as close as possible to a radial direction. To this end, we minimise the angle $\beta$, between $\vec{r_s}$ and $\vec{b}$ such that the projection of $\vec{b}$ yields $\vec{r_{h}}$. Consequently, $\beta$ is the minimum angle, an incompressible lower limit, that can exist between the radial vector and the major axis vector of a dwarf galaxy according to our data. The intrinsic ellipticity (assuming this ``radial'' hypothesis) $E_{R_{ki}}$ is thus obtained (see Figure~\ref{Rad}) coupled with $\beta_{ki}$ (see Figure~\ref{beta}). For more details, see appendix~\ref{annexeradial}.

\begin{figure}
\begin{center}
\includegraphics[scale=0.32]{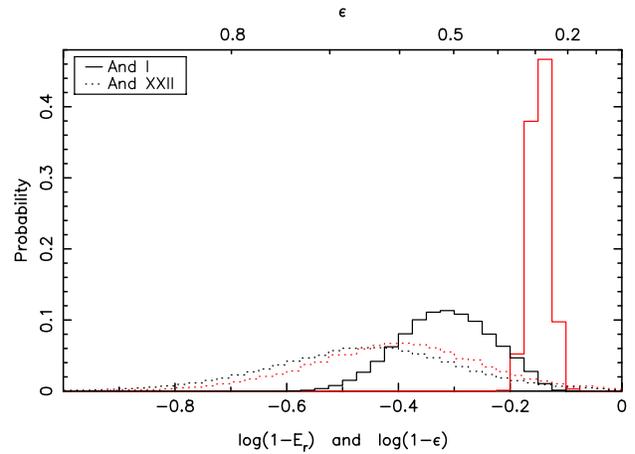}
\caption{\small\label{Rad}
Histogram of the ellipticity for And I (solid black line) and And XXII (dotted black line) after radial deprojection obtained from $10^5$ draws of the MCMC chain. The red curves are the observed ellipticity previously seen in Figure~\ref{Nico}.}
\end{center}
\end{figure}

\begin{figure}
\begin{center}
\includegraphics[scale=0.32]{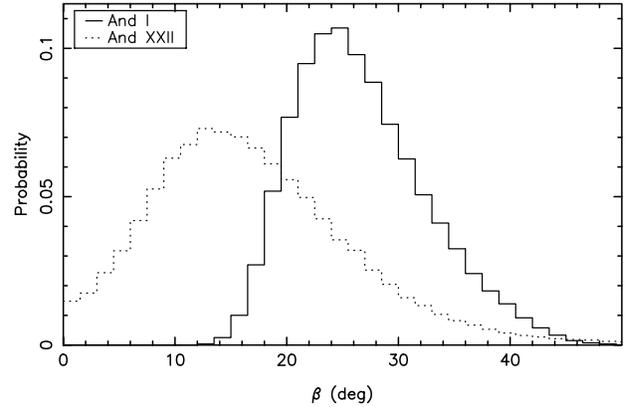}
\caption{\small\label{beta}
Probability density function of the angle $\beta$ between the major axis of : And I (solid line), And XXII (dotted line) and their respective radial direction after radial deprojection (obtained from $10^5$ MCMC draws).}
\end{center}
\end{figure}

\begin{figure}
\begin{center}
\vbox{
\includegraphics[scale=0.32]{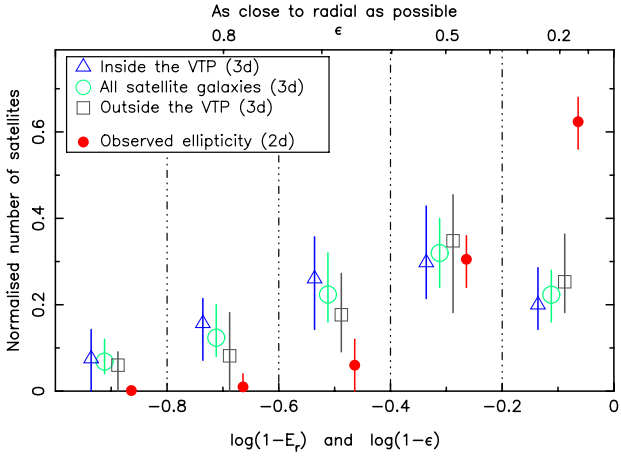}
}
\caption{\small\label{all_rad}
As Figure~\ref{all_tang}, but for the radial deprojection method.}
\end{center}
\end{figure}

The radially deprojected ellipticity is displayed in Figure~\ref{all_rad}, and interestingly, the roundest bin is not dominant, and a significant number of highly elliptical systems are required if this assumption is realistic. Again, differences between galaxies within the VTP of galaxies and outside of it are not significant.

\subsection{$\gamma$-deprojection (random angle within $\cal{P}$)}

\begin{figure}
\begin{center}
\includegraphics[scale=0.32]{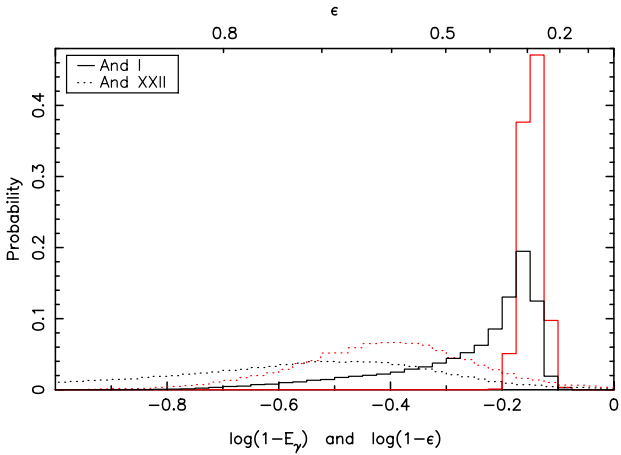}
\caption{\small\label{gamDeprojI}
Histogram of the ellipticity for And I (solid black line) and And XXII (dotted black line) after $\gamma$-deprojection obtained from $10^5$ draws of the MCMC chain. The red curves are the observed ellipticity previously seen in Figure~\ref{Nico}.}
\end{center}
\end{figure}

\begin{figure}
\begin{center}
\includegraphics[scale=0.32]{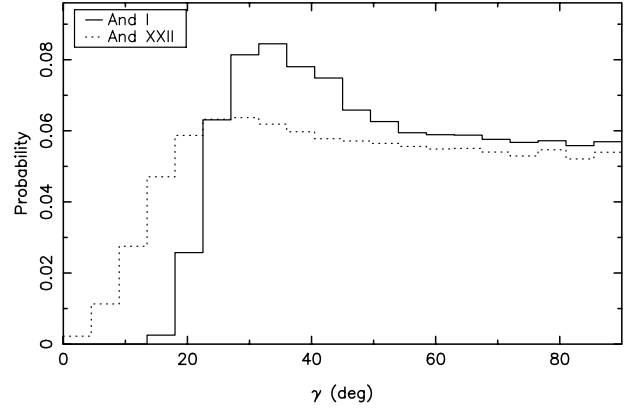}
\caption{\small\label{gammaAngle}
Probability density function of the angle $\gamma$ (in degrees) between the major axis of And I (solid line) and And XXII (dotted line) with their respective radial direction after $\gamma$-deprojection (obtained from $10^5$ MCMC draws).}
\end{center}
\end{figure}

In the previous subsections we considered a ``tangential'' and a ``radial'' de-projection. We have seen that a satellite can always be fit to a 3-D tangential orientation, but that a prolate system cannot be placed closer to a radial direction than a minimum angle $\beta$ (which depends on the structure and position of the satellite). However in reality, we do not have observational information that allows us to favor either of these orientations, and we only know that the major axis lies within the line of sight plane $\cal{P}$.
This motivates our final deprojection, where we instead make the minimal assumption that the satellites are randomly oriented within $\cal{P}$. The random angle $\gamma$, is the angle between the line of sight and the major axis $\vec{b}$. The intrinsic ellipticity (assuming this ``gamma'' hypothesis) $E_{\gamma_{ki}}$ is thus obtained (see Figure~\ref{gamDeprojI}) coupled with $\gamma_{ki}$ (see Figure~\ref{gammaAngle}).

\begin{figure}
\begin{center}
\includegraphics[scale=0.32]{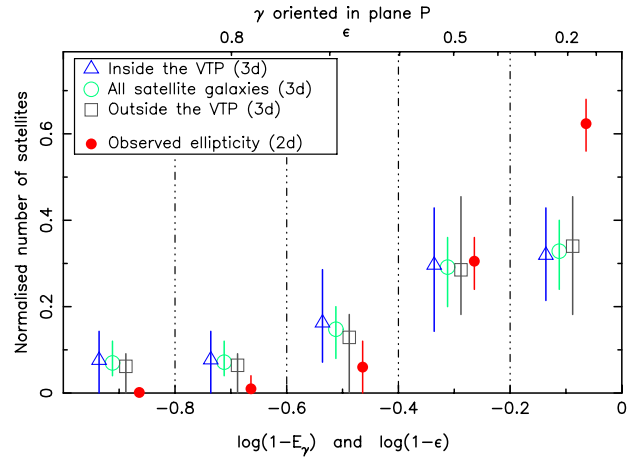}
\caption{\small\label{gamDeproj}
As Figure~\ref{all_tang}, but for the $\gamma$ deprojection method (where we assume that the major axis lies in plane $\cal{P}$ at a random angle with respect to the line of sight).}
\end{center}
\end{figure}

The resulting deprojection is shown in Figure~\ref{gamDeproj} and presents a rather flat ellipticity distribution. Again, there is almost no difference between the dwarf galaxies that belong to the VTP and those outside of it.

\section{Re-projection tests}\label{results}

The distributions of ellipticity obtained in Figures~\ref{all_tang} and~\ref{all_rad} take into account the observational information with a hypothesis on the orientation and on the prolate shape of the satellite galaxies. In contrast, the distributions of ellipticity obtained in Figure~\ref{gamDeproj} take into account the observational information assuming that the shape of the satellite galaxies is prolate, but that no particular direction can be privileged. If either of these approaches is realistic, one of the resulting deprojected distributions should be the real intrinsic ellipticity. 

To test the consistency of this approach, we decided to examine whether the derived deprojected ellipticity distributions would give rise to projected distributions similar to the observed distribution of $\epsilon$ for random viewing directions.

\begin{figure}
\begin{center}
\includegraphics[scale=0.32]{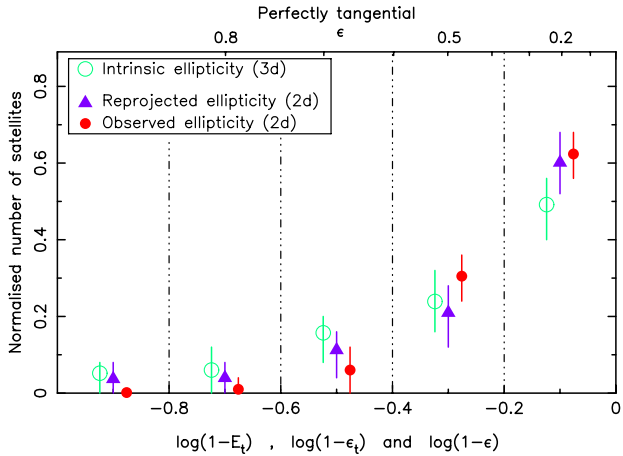}
\caption{\small\label{tan_reproj}
Distribution of projected ellipticity for 1000 sets of 25 dSph satellites randomly built with intrinsic ellipticity picked from the tangentially deprojected distribution (purple triangles). Red dots show the distribution of observed ellipticities. Green circles are the intrinsic ellipticity determined with the tangential deprojection method represented in Figure~\ref{all_tang}. Vertical error bars indicate one standard deviation. }
\end{center}
\end{figure}

\begin{figure}
\begin{center}
\includegraphics[scale=0.32]{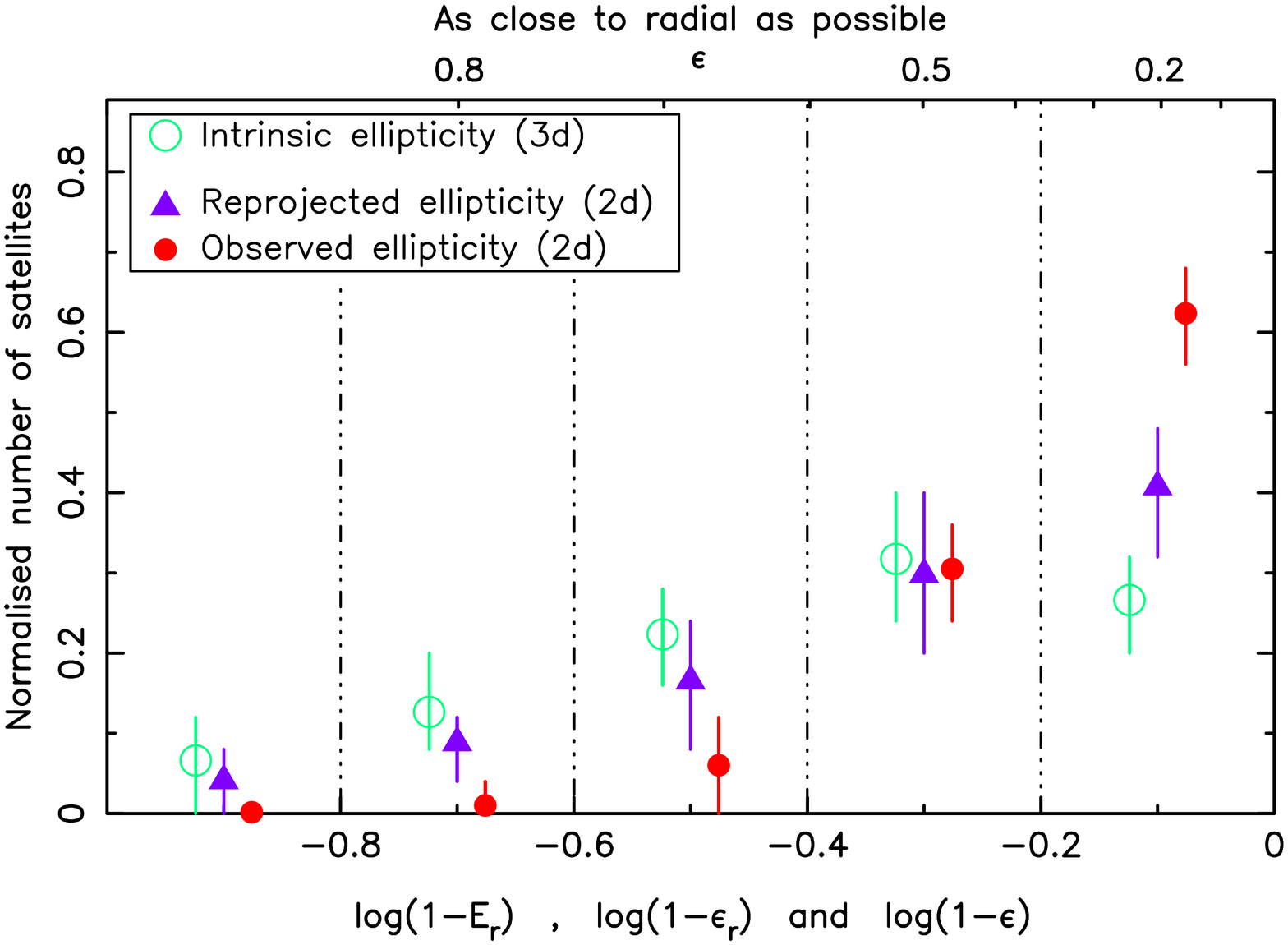}
\caption{\small\label{rad_reproj}
As Figure~\ref{tan_reproj}, but for the radial deprojection method.}
\end{center}
\end{figure}

\begin{figure}
\begin{center}
\includegraphics[scale=0.32]{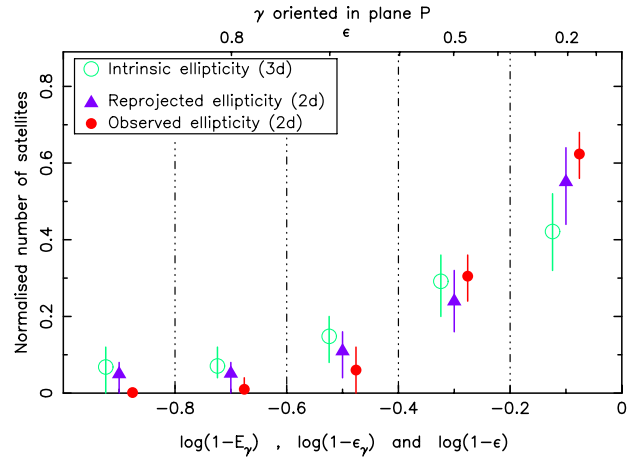}
\caption{\small\label{gam_reproj}
As Figure~\ref{tan_reproj}, but for the $\gamma$-deprojection method.}
\end{center}
\end{figure}

To this end we modelled systems containing 25 (3-dimensional) prolate ellipsoids with random orientation. The ellipticity, and so the semimajor axis, is randomly picked from the 3-D tangential-deprojected ellipticity distribution. The positions of the dwarf galaxy models are drawn randomly inside a sphere of $450\kpc$ around the host. A set of 1000 such systems were generated and observed from a position equivalent to that of the Earth point at $780\kpc$ from the host galaxy. The final 2-D projected ellipticity is shown Figure~\ref{tan_reproj}. The same test is applied with the 3-D radial-deprojected ellipticity and the 3-D $\gamma$-deprojected one. Results are shown Figures~\ref{rad_reproj} and~\ref{gam_reproj}, respectively.

In Figure~\ref{rad_reproj} one can see that the projected ellipticity distribution (purple triangles) based on the radial deprojection does not reproduce well the observed distribution of ellipticity (red circles) in the roundest bin. This means that if dwarf galaxies are prolate structures, and if our viewing direction is not special, the galaxies are not oriented in a radial configuration. However, the projected distributions based on tangential (Figure~\ref{tan_reproj}) or the $\gamma$-deprojection (Figure~\ref{gam_reproj}) methods are perfectly consistent with the observed distribution of ellipticity. We consider that this is a good consistency test of our method. Thus modelling dwarf galaxies as possessing prolate shapes can give self-consistent solutions in terms of their derived ellipticity distribution. Furthermore, the population of dwarf galaxies around Andromeda can be considered as a drawing from a distribution of general intrinsic ellipticity.

\section{Individual cases and discussion}
\label{individual}

The results for all of the dSph galaxies are listed in table~\ref{tabIndiv}. The second column contains the character "p" if the galaxy belongs to the VTP according \cite{Ibata13}, and "a" otherwise. Columns 3 and 4 contain the mean observed ellipticity and the mean observed half-light radius obtained with the method of \cite{Martin08}. Column 5 is the mean distance from M31 to the centre of the dSph galaxy obtained by \cite{Conn12}. Note that these observed average values (columns 3, 4 and 5) take into account the fact that some values have been rejected by our tidal criterion in the $\gamma$-deprojection method. Columns 6, 8 and 10 represent respectively the mean of the tangential-deprojected, radial-deprojected and $\gamma$-deprojected ellipticity obtained with our method. Columns 7, 9 and 11 are the percentage of excluded cases by our tidal criteria for the 3 deprojections. The last column is the angle $\beta$ previously defined as the geometrically minimum possible angle between the radial direction and the plane $\cal{P}$. All uncertainties indicate one standard deviation of each parameter with our method.

\begin{table*}
 \centering
  \caption{Properties of the 25 satellites of Andromeda studied here, showing the results of the 3 different deprojections obtained from $10^5$ draws of the MCMC chain. (1) \citet{Ibata13} (2) According to \citet{Crnojevic14} for NGC147 and NGC185 and to Martin et al. (2015) (in preparation) for the other galaxies (3) \citet{Conn12} * Note that these values are calculated according to the $\gamma$-deprojection tidal rejection criterion.} \label{tabIndiv}
  \begin{tabular}{@{}lrrrrrrrrrrr@{}}
  \hline
   Name & VTP & $\epsilon$ & r$_h$     & $d_{M31}$    & E$_t$ & Tide & E$_r$ & Tide  & E$_\gamma$ & Tide     & $\beta$ \\
               &          &                    &(arcmin)  & (\kpc)              &            & (\%)   &             &   (\%) &                          &     (\%)  &    (deg) \\
               &  (1)   &    (2)*           &   (2)*         &   (3)*                 &            &           &             &            &                          &              &               \\
 \hline
AndI       & p & $          0.29{\pm          0.03}$ & $          3.97{\pm          0.14}$ & $          68.7{\pm          13.3}$ & $          0.37{\pm          0.05}$ & $           1.8$ & $          0.51{\pm          0.10}$ & $           0.0$ & $          0.43{\pm          0.15}$ & $          14.3$ & $          26.7{\pm           6.0}$\\
AndII      & a & $          0.14{\pm          0.02}$ & $          5.13{\pm          0.10}$ & $         197.0{\pm          11.0}$ & $          0.15{\pm          0.02}$ & $           0.0$ & $          0.57{\pm          0.09}$ & $           0.3$ & $          0.31{\pm          0.20}$ & $           5.0$ & $          43.4{\pm           3.3}$\\
AndIII     & p & $          0.59{\pm          0.03}$ & $          1.89{\pm          0.15}$ & $          91.2{\pm          13.9}$ & $          0.66{\pm          0.05}$ & $           0.4$ & $          0.76{\pm          0.06}$ & $           0.0$ & $          0.71{\pm          0.13}$ & $           6.8$ & $          40.9{\pm           7.7}$\\
AndV       & a & $          0.28{\pm          0.07}$ & $          1.68{\pm          0.14}$ & $         116.3{\pm           5.9}$ & $          0.70{\pm          0.15}$ & $           7.9$ & $          0.31{\pm          0.07}$ & $           0.0$ & $          0.47{\pm          0.22}$ & $           3.3$ & $          16.2{\pm           7.1}$\\
AndIX      & p & $          0.12{\pm          0.07}$ & $          1.85{\pm          0.21}$ & $         138.4{\pm          52.3}$ & $          0.12{\pm          0.08}$ & $           0.0$ & $          0.63{\pm          0.23}$ & $           6.5$ & $          0.29{\pm          0.23}$ & $           1.9$ & $          11.0{\pm           7.2}$\\
AndX       & a & $          0.30{\pm          0.18}$ & $          1.15{\pm          0.24}$ & $         143.9{\pm          27.6}$ & $          0.36{\pm          0.20}$ & $           0.1$ & $          0.54{\pm          0.19}$ & $           0.4$ & $          0.47{\pm          0.28}$ & $           2.6$ & $          12.5{\pm           8.9}$\\
AndXI      & p & $          0.24{\pm          0.16}$ & $          0.78{\pm          0.20}$ & $         141.0{\pm          41.7}$ & $          0.46{\pm          0.31}$ & $           2.6$ & $          0.41{\pm          0.27}$ & $           0.4$ & $          0.43{\pm          0.29}$ & $           1.3$ & $          34.4{\pm          24.1}$\\
AndXII     & p & $          0.39{\pm          0.25}$ & $          2.85{\pm          1.36}$ & $         157.7{\pm          52.2}$ & $          0.47{\pm          0.26}$ & $          15.2$ & $          0.56{\pm          0.24}$ & $           5.1$ & $          0.50{\pm          0.27}$ & $          18.7$ & $          15.9{\pm          12.9}$\\
AndXIII    & p & $          0.54{\pm          0.16}$ & $          1.11{\pm          0.41}$ & $         180.2{\pm          81.2}$ & $          0.71{\pm          0.19}$ & $           6.9$ & $          0.65{\pm          0.17}$ & $           0.0$ & $          0.68{\pm          0.19}$ & $           3.1$ & $           9.8{\pm           7.4}$\\
AndXIV     & p & $          0.20{\pm          0.11}$ & $          1.55{\pm          0.16}$ & $         184.4{\pm          34.9}$ & $          0.58{\pm          0.28}$ & $          10.1$ & $          0.25{\pm          0.13}$ & $           0.0$ & $          0.39{\pm          0.25}$ & $           3.0$ & $          13.4{\pm          10.4}$\\
AndXV      & a & $          0.24{\pm          0.10}$ & $          1.39{\pm          0.13}$ & $         167.9{\pm          45.5}$ & $          0.26{\pm          0.10}$ & $           0.4$ & $          0.73{\pm          0.20}$ & $          18.7$ & $          0.42{\pm          0.24}$ & $           3.5$ & $          36.2{\pm          13.7}$\\
AndXVI     & p & $          0.29{\pm          0.08}$ & $          1.00{\pm          0.08}$ & $         316.1{\pm          33.7}$ & $          0.30{\pm          0.08}$ & $           0.0$ & $          0.79{\pm          0.08}$ & $           0.2$ & $          0.49{\pm          0.23}$ & $           0.7$ & $          21.9{\pm           3.0}$\\
AndXVII    & p & $          0.42{\pm          0.10}$ & $          1.39{\pm          0.21}$ & $          71.8{\pm          21.0}$ & $          0.49{\pm          0.12}$ & $           7.4$ & $          0.66{\pm          0.15}$ & $           0.5$ & $          0.55{\pm          0.16}$ & $          13.8$ & $          30.5{\pm          11.0}$\\
AndXVIII   & a & $          0.21{\pm          0.13}$ & $          0.78{\pm          0.11}$ & $         454.6{\pm          40.1}$ & $          0.22{\pm          0.14}$ & $           0.0$ & $          0.73{\pm          0.14}$ & $           0.4$ & $          0.40{\pm          0.28}$ & $           0.4$ & $           7.6{\pm           4.8}$\\
AndXIX     & a & $          0.41{\pm          0.09}$ & $         11.65{\pm          1.64}$ & $         183.8{\pm          55.2}$ & $          0.49{\pm          0.08}$ & $          74.4$ & $          0.56{\pm          0.17}$ & $          65.2$ & $          0.49{\pm          0.12}$ & $          72.5$ & $           3.7{\pm           2.7}$\\
AndXX      & a & $          0.30{\pm          0.20}$ & $          0.77{\pm          0.25}$ & $         141.7{\pm          13.4}$ & $          0.57{\pm          0.31}$ & $           2.9$ & $          0.38{\pm          0.24}$ & $           0.2$ & $          0.48{\pm          0.30}$ & $           1.5$ & $          32.5{\pm          24.9}$\\
AndXXI     & a & $          0.32{\pm          0.11}$ & $          4.18{\pm          0.61}$ & $         137.2{\pm           9.7}$ & $          0.54{\pm          0.14}$ & $          12.9$ & $          0.39{\pm          0.12}$ & $           0.1$ & $          0.44{\pm          0.17}$ & $          16.9$ & $          36.2{\pm          10.8}$\\
AndXXII    & a & $          0.58{\pm          0.14}$ & $          1.14{\pm          0.32}$ & $         261.5{\pm          33.6}$ & $          0.77{\pm          0.13}$ & $           3.3$ & $          0.63{\pm          0.13}$ & $           0.0$ & $          0.72{\pm          0.17}$ & $           3.2$ & $          17.1{\pm           8.6}$\\
AndXXIII   & a & $          0.38{\pm          0.05}$ & $          5.43{\pm          0.40}$ & $         130.3{\pm           3.4}$ & $          0.72{\pm          0.07}$ & $          57.4$ & $          0.40{\pm          0.06}$ & $           0.0$ & $          0.51{\pm          0.14}$ & $          16.1$ & $          35.3{\pm           4.7}$\\
AndXXIV    & a & $          0.23{\pm          0.15}$ & $          2.92{\pm          0.72}$ & $         166.7{\pm          24.8}$ & $          0.30{\pm          0.19}$ & $           0.9$ & $          0.45{\pm          0.24}$ & $           2.9$ & $          0.38{\pm          0.26}$ & $           6.8$ & $          25.8{\pm          13.4}$\\
AndXXV     & p & $          0.20{\pm          0.11}$ & $          3.18{\pm          0.33}$ & $         111.9{\pm          32.2}$ & $          0.33{\pm          0.17}$ & $          21.2$ & $          0.32{\pm          0.17}$ & $           1.3$ & $          0.31{\pm          0.18}$ & $          18.3$ & $          23.3{\pm          14.0}$\\
AndXXVI    & p & $          0.30{\pm          0.20}$ & $          1.45{\pm          0.49}$ & $         197.6{\pm         108.9}$ & $          0.43{\pm          0.27}$ & $           2.7$ & $          0.54{\pm          0.28}$ & $           0.8$ & $          0.48{\pm          0.29}$ & $           2.1$ & $          20.6{\pm          17.0}$\\
CasII     & p & $          0.38{\pm          0.07}$ & $          1.47{\pm          0.14}$ & $         176.2{\pm          40.4}$ & $          0.45{\pm          0.09}$ & $           0.3$ & $          0.69{\pm          0.12}$ & $           0.0$ & $          0.57{\pm          0.21}$ & $           1.1$ & $          36.1{\pm          10.0}$\\
NGC147     & p & $          0.46{\pm          0.02}$ & $          6.70{\pm          0.09}$ & $         118.6{\pm          10.1}$ & $          0.63{\pm          0.07}$ & $           1.6$ & $          0.58{\pm          0.06}$ & $           0.0$ & $          0.60{\pm          0.15}$ & $          10.4$ & $          39.1{\pm           4.8}$\\
NGC185     & p & $          0.22{\pm          0.01}$ & $          2.94{\pm          0.04}$ & $         181.9{\pm          15.7}$ & $          0.24{\pm          0.01}$ & $           0.0$ & $          0.60{\pm          0.04}$ & $           0.0$ & $          0.42{\pm          0.23}$ & $           1.2$ & $          24.8{\pm           2.3}$\\

  \hline
\end{tabular}
\end{table*}

First, the most powerful geometric constraint is a high angle $\beta$. Indeed, it corresponds to the lower limit for a radial orientation. This means that even if the orientation is unknown, a dwarf galaxy cannot be closer to the radial direction than $\beta$ (see Figure 1). This criterion does not depend at all on our hypothesis. Thus, in the present study, we have identified 6 dwarf galaxies - And II, And III, And XXI, And XXIII, CasII and NGC 147, which can absolutely not be radially oriented. For this we have chosen a criterion such that $\beta - 1\sigma > 25\deg$ (where $\sigma$ is here the standard deviation of $\beta$). These dwarf galaxies are relatively close to their host. Moreover, they mostly have a large half-light radius and favour a high ellipticity (more than 0.5) in each kind of deprojection. Also, for 5 of these 6 galaxies, the $\gamma$-deprojection shows the largest rate of tidal rejections among the 3 cases. These particularities suggest that these six dwarf galaxies may be largely affected by their environment and could be affected by tides or even remnants of ancient tidal dwarf galaxies (TDG). For the most massive object, NGC 147, This last result is in a good agreement with \cite{Crnojevic14} where they have recently found that NGC 147 presents tidal tails. Moreover, it could also indicate that it is beginning to be stretched due to gravitational interactions, as attested by the stellar stream that emanates from it \citep{McConnachie09}.

It is also interesting to see the deprojection behaviour of the other dwarf galaxies that have an observed large half-light radius and a high level of tidal rejections with all 3 deprojection methods. And XII, And XIX and And XXV are such cases. This might be due to the effects of tidal interactions with M31, as argued by \cite{Collins14} for And XIX and And XXV (as well as And XXI). Moreover, this strongly suggests that these galaxies are close to being destroyed and becoming stellar streams. Please note that we have been conservative in estimating the tidal limits of the dwarf galaxies, as we used the measured velocity dispersion {\it plus} one standard deviation to calculate the tidal limits. Thus the percentage of tidal rejections listed in Table~\ref{tabIndiv} should be considered as lower limits.

Finally, And XIII and And XXII have a large intrinsic ellipticity with all of the deprojection choices. Indeed, with the 3 deprojections, they always conserve an ellipticity around 0.7. Thus, these dSph galaxies are intrinsically very elongated.

\begin{figure}
\begin{center}
\includegraphics[scale=0.32]{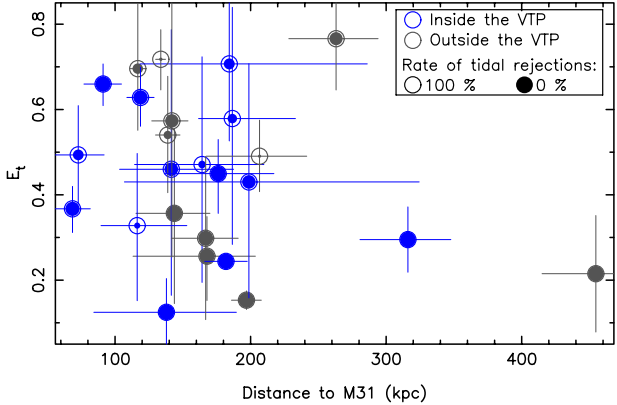}
\caption{\small\label{ellvsdist}
Tangentially deprojected ellipticity versus distance to Andromeda. Blue circles are the values for the satellites in the VTP of galaxies~\citep{Ibata13} while grey circles are for the ones outside. The filling of these circles represents the tidal rejection rate (see labels in the top-right box).}
\end{center}
\end{figure}

As the re-projection tests suggest a possible tangential orientation being very plausible, we have examined in this case the deprojected ellipticity as a function of the distance (Figure~\ref{ellvsdist}). Blue circles are the values for the satellites within the VTP and grey circles are the values for those outside. There is a priori no obvious correlation. Nevertheless, a close inspection of the repartition of the satellites outside of the VTP seems to indicate a relatively tight correlation between 100 and 200 kpc : dSphs are more elongated when closer to M31. No such correlation is present for satellites in the VTP. With the large error bars, no strong conclusion can be drawn, but it would be interesting to compare this with different models, both in standard cosmology in which the VTP could have been recently accreted and in alternative scenarios such as those based on TDGs.

We next examine the properties of the population of 25 dwarf galaxies as a whole, which we consider as a representative sample for this kind of object. The ellipticity of this group is listed in Table~\ref{tabTot} for projected ($\epsilon$), tangential-deprojected (E$_t$), radial-deprojected (E$_r$) and $\gamma$-deprojected (E$_{\gamma}$) with their total percentage of tidal rejection.
\begin{table}
 \centering
  \caption{Average ellipticities of the whole sample of dwarf galaxies of Andromeda used in this work, obtained from $25 \times 10^5$ draws of the MCMC chain.}\label{tabTot}
  \begin{tabular}{@{}lrr@{}}
  \hline
                   & Mean ellipticity & Tidally rejected  \\
                   &                             &  (\%)                      \\
 \hline
$\epsilon$ & $          0.32{\pm          0.11}$ & $          -- $\\
E$_t$ & $          0.44{\pm          0.14}$ & $          9.22$\\
E$_r$ & $          0.54{\pm          0.15}$ & $          4.12$\\
E$_{\gamma}$ & $          0.48{\pm          0.21}$ & $          9.14$\\
  \hline
\end{tabular}
\end{table}
This percentage is compatible with a physical behaviour with a high value for tangential and $\gamma$ deprojections and a much lower value for the radial deprojection. Objects in a radial configuration are falling toward their host with a compact shape whereas objets in tangential or random orientation are suspected to have been strongly disrupted in their orbits around their host.

As expected, each kind of deprojection gives a larger ellipticity (with a minimum of 0.44) than the projected observations. Thus, these dwarf galaxies tend to possess an elongated shape with a minor to major axis ratio ($a/b$) of approximately 1/2. \cite{VeraCiro14} and \cite{Barber15} find a sightly larger but coherent axis ratio in the inner regions ($\sim 1 \kpc$) of the dark matter satellites in the Aquarius simulation that could possibly host stars. Thus our conclusion appears to be consistent with the possibility that stellar components in the Andromeda satellite galaxies possess the same shape. This is perhaps not incompatible either with them being remnants of ancient tidal dwarf galaxies. These observational values can in any case serve as a test of various models in the future.

\section{Conclusions}\label{conclusion}

First of all, we have seen that adopting a prolate shape for the dwarf satellite galaxies of Andromeda gives a coherent solution in the sense that the derived ellipticity distribution for the real viewing direction is similar to what would be derived from a random vantage point. 
This simplifying assumption has allowed us to derive the intrinsic ellipticity for the entire population of satellites. The resulting ellipticity distribution is found to be relatively flat between $log(1-\epsilon)$ of $0$ to $-0.4$ before decaying at larger ellipticities. Some of the dwarf galaxies are found to be intrinsically significantly elongated. This is probably not incompatible with them being remnants of ancient tidal dwarf galaxies as recently found by, e.g., \citet{Yang14}, but it could also be consistent with dwarf galaxies having the same shape as their hosting dark matter subhalo. Secondly, no clear difference in ellipticity between the satellites inside the VTP and those outside of it was noticed. The two populations of galaxies seem to be made of the same type of objects as also recently concluded by \cite{Collins15}. Nevertheless we noted that the distribution of ellipticity versus distance to M31 in the tangential deprojection case was slightly different for galaxies inside and outside of the VTP.

\appendix

\section[]{Tangential case}\label{annexetangent}
We seek the vector $\vec{m}$ which has the same projection on the sky as the half-light radius $\vec{r_h}$. This means it has the same projected position angle (PA) (see eq.~\ref{pa}). At the same time, $\vec{m}$ must be perpendicular to the direction $\vec{r_s}$, between M31 and the centre of the dSph (see eq.~\ref{tang}). First of all, for a dSph, we define a local coordinate system ($\vec{los}$,$\vec{n}$,$\vec{e}$) centred on the galaxy where the line of sight $\vec{los}$ is one of the axes and where $\vec{n}$ points towards the north. From (standard) sky coordinates ($\chi$,$\eta$) with respect to the  centre, we obtain the local base:
\begin{equation}\label{base}
\vec{los} = \begin{pmatrix} \cos \eta \sin \chi \\  \sin \eta \\ \cos \eta \cos \chi \end{pmatrix}, \vec{n} = \begin{pmatrix} -\sin \chi \sin \eta \\  \cos \eta \\ -\cos \chi \sin \eta\end{pmatrix}, \vec{e} = \begin{pmatrix} \cos \chi \\  0 \\ -\sin \chi \end{pmatrix}
\end{equation}
From this base with our conditions and the normalisation of $\vec{m}$, a 3 equation system can be deduced:
\begin{equation}\label{tang}
\vec{r_s} . \vec{m} = 0
\end{equation}
\begin{equation}\label{pa}
\tan (PA) = \frac{\vec{m}.\vec{e}}{\vec{m}.\vec{n}}
\end{equation}
\begin{equation}\label{norm}
{m_x}^2 + {m_y}^2 + {m_z}^2 = 1  \, .
\end{equation}
The solution of this system gives us:
\begin{equation}\label{m}
\vec{m} = \begin{pmatrix} \frac{-1}{r_{s_x}\sqrt{C}}(\frac{Ar_{s_y}}{B}+r_{s_z})  \\ \frac{A}{B\sqrt{C}} \\ \frac{1}{\sqrt{C}} \end{pmatrix} \\
\end{equation}
where:
\begin{align*}
\begin{cases}
A& = \tan (PA) \cos \chi \sin \eta r_{s_x} - \tan(PA) r_{s_z} \sin \chi \sin \eta \\
& - r_{s_z} \cos \chi - r_{s_x} \sin \chi \\
B& = \tan(PA)r_{s_y}\sin \chi \sin \eta + r_{s_x}\tan(PA)\cos\eta \\
& + r_{s_y}\cos\chi \\
C& = \frac{1}{r_{s_x}^2}\left( \frac{Ar_{s_y}}{B} + r_{s_z}\right)^2 + \left( \frac{A}{B}\right)^2 + 1
\end{cases}
\end{align*}
As we know $\vec{m}$, the norm of the tangentially deprojected semimajor axis $b_t$ has yet to be discovered. DSph are considered as prolate (see eq.~\ref{prolate}). Thus, we are putting ourselves this time in the plane defined by $\vec{los}$ and $\vec{m}$. The new local base ($\vec{y}$,$\vec{z}$) corresponds to the minor (a) and the semimajor axis ($b_t$) of the real prolate.
\begin{equation}\label{prolate}
\frac{y^2}{a^2} + \frac{z^2}{{b_t}^2} = 1
\end{equation}
For the present purposes, dSph around M31 are far enough to be considered as being at infinity. Consequently, $r_h$ is approximately seen between two parallel rays, $\vec{los}$ and $\vec{los'}$ with a separation of $\omega$. We called $\alpha$ the angle between $\vec{y}$ and $\vec{los}$. $y_0$ is the point at the intersection between $\vec{y}$ and $\vec{los'}$. It follows that:
\begin{equation}\label{sin}
\sin{\alpha} = \frac{\omega}{y_0}
\end{equation}
From eq.~\ref{prolate}, we obtain y and:
\begin{equation}\label{pente}
\frac{dy}{dz} = \frac{-az}{{b_t}^2 \sqrt{1-\frac{z^2}{{b_t}^2}}}
\end{equation}
With eq.~\ref{pente}, we obtain the equation for the straight line $(d)$ which has $\vec{los'}$ for guiding vector:
\begin{equation}\label{droite}
y = y_0 - \frac{az^2}{{b_t}^2\sqrt{1-\frac{z^2}{{b_t}^2}}}
\end{equation}
And so, the intersection between $(d)$ (eq.~\ref{droite}) and the prolate (eq.~\ref{prolate}) with eq.~\ref{sin} gives:
\begin{equation}\label{inter}
a = \frac{\omega}{\sin{\alpha}}\sqrt{1-\frac{z^2}{{b_t}^2}}
\end{equation}
The development of eq.~\ref{inter} with $\tan{\alpha} = -y'$ and $\beta_t = \frac{\pi}{2} - \alpha$ gives the final real semimajor axis, tangentially deprojected:
\begin{equation}\label{inter}
b_t = \sqrt{\left(\frac{\omega}{\sin{\beta_t}}\right)^2-\left(\frac{a}{\tan{\beta_t}}\right)^2}
\end{equation}

\section[]{Radial case}\label{annexeradial}
We seek the vector $\vec{q}$ which has the same projection on the sky as the half-light radius $\vec{r_h}$. It means the same projected position angle (PA). At the same time, $\vec{q}$ must be as close as possible to the radial direction $\vec{r_s}$, connecting M31 with the centre of the dSph (see eq.~\ref{tang}). We place the system in the same basis as the tangential deprojection (see appendix~\ref{annexetangent}). The line of sight is again $\vec{los}$ and $\vec{m}$ is the tangential direction. Thus, $\vec{q}$ must be placed in the plane $\mathcal{P}$ defined by these two vectors. $\mathcal{P}$ has as its normal vector $\vec{n_\mathcal{P}}$:
\begin{equation}\label{normal}
\vec{n_\mathcal{P}} = \begin{pmatrix} Bn_{\mathcal{P}_z}  \\ An_{\mathcal{P}_z} \\ \frac{1}{\sqrt{1+A^2+B^2}} \end{pmatrix} \\
\end{equation}
where:
\begin{align*}
\begin{cases}
A& =  \frac{r_xm_z-m_xr_z}{-r_xm_y+m_xr_y} \\
B& =  \frac{-m_y}{m_x}A - \frac{m_z}{m_x}   \\
\end{cases}
\end{align*}
And so, with the minimum angle between $\vec{q}$ and $\vec{r_s}$, the constraint of lying in Plane $\mathcal{P}$ and the normalisation of $\vec{q}$, a system of 3 equations can be deduced:
\begin{align}\label{radial}
\begin{cases}
\vec{q}.\vec{n_\mathcal{P}}& =  0 \\
\vec{q}.\vec{r_s}& =  max   \\
q_x^2+q_y^2+q_z^2& = 1 \\
\end{cases}
\end{align}
This system is solved by maximising eq.~\ref{radial}:
\begin{equation}\label{q}
\vec{q} = \begin{pmatrix} \frac{-1}{n_{\mathcal{P}_x}}(n_{\mathcal{P}_x} q_y+n_{\mathcal{P}_z} q_z) \\ \frac{-4n_{\mathcal{P}_y} n_{\mathcal{P}_z}}{n_{\mathcal{P}_y}^2+n_{\mathcal{P}_x}^2}q_z \\ \frac{A\sqrt{1-n_{\mathcal{P}_z}^2}}{\sqrt{A^2+B^2}} \end{pmatrix} \\
\end{equation}
where:
\begin{align*}
\begin{cases}
A& =  r_{s_z}((n_{\mathcal{P}_x}^2+n_{\mathcal{P}_y}^2) - n_{\mathcal{P}_y}n_{\mathcal{P}_z}) - r_{s_x}n_{\mathcal{P}_z}n_{\mathcal{P}_x}\\
B& = n_{\mathcal{P}_x}r_{s_y} - n_{\mathcal{P}_y}r_{s_x}  \\
\end{cases}
\end{align*}
Finding the norm $b_r$ follows the same approach as that for the tangential case and so:
\begin{equation}\label{intera}
b_r = \sqrt{\left(\frac{\omega}{\sin{\beta_r}}\right)^2-\left(\frac{a}{\tan{\beta_r}}\right)^2}
\end{equation}

\label{lastpage}

\end{document}